\documentclass[12pt,preprint]{aastex}
\usepackage{graphicx}

\usepackage{natbib}

\newcommand{\be}{\begin{equation}}
\newcommand{\ee}{\end{equation}}
\newcommand{\nn}{\mbox{} \nonumber \\ \mbox{} }
\newcommand{\ba}{\begin{eqnarray}}
\newcommand{\ea}{\end{eqnarray}}
\newcommand{\om}{\omega}
\newcommand{\Alfven}{ Alfv\'{e}n }

\newcommand{\E}{{\bf E}}
\newcommand{\B}{{\bf B}}

\newcommand{\Bf}{{magnetic field}}
\newcommand{\Bfs}{{magnetic fields}}
\newcommand{\Ef}{{electric  field\,}}
\newcommand{\Efs}{{electric fields\,}}
\newcommand{\NS}{neutron star}
\newcommand{\ms}{magnetosphere}
\newcommand{\NSs}{{neutron stars}}

\newcommand{\mss}{magnetospheres}
\newcommand{\Lf}{Lorentz factor}

\newcommand{\EM}{electromagnetic}

\newcommand\eg{\textit{e.g.}}

\newcommand\lo{\mathrel{\raise.3ex\hbox{$<$}\mkern-14mu\lower0.6ex\hbox{$\sim$}}}
\newcommand\go{\mathrel{\raise.3ex\hbox{$>$}\mkern-14mu\lower0.6ex\hbox{$\sim$}}}

\begin{document}
\title{Free electron laser in magnetars/Fast Radio Bursts}

\author{Maxim Lyutikov}
\author{
Department of Physics and Astronomy, Purdue University, 
 525 Northwestern Avenue,
West Lafayette, IN
47907-2036 }

\begin{abstract}
We discuss  coherent  free electron laser (FEL)   operating during explosive reconnection events in magnetized pair  plasma of magnetar magnetospheres. The model explains  many salient features of  Fast Radio Bursts/magnetars' radio emission: temporal coincidence of radio and high energy bursts,   high efficiency of conversion of plasma kinetic energy into coherent radiation,  presence of variable, narrow-band emission features drifting down in frequency,   high degree of linear polarization. 
The model relies on magnetar-specific drifting $e^\pm$ plasma components (which generate wiggler field due to the development of  the firehose instability)  and the presence   
of reconnection-generated particle beam with   mild  \Lf\ of $\gamma_b\sim$ few hundred. 
  \end{abstract}

\maketitle

\section{Introduction}

Generation of high brightness coherent emission by various types of \NSs\ is a major unresolved problem in astrophysics  -- for nearly fifty years. 
One of the difficulty in identifying the process is that  radio waves carry minuscule relative  amount of the  energy, $\sim 10^{-5}$ is typical. 
The phenomenon of Fast Radio Bursts challenges our understating of relativistic plasma coherent processes to the extreme. In this case radio waves can indeed carry astrophysically important amount of the energy (\eg, radio luminosity can match, for a short period of time, the macroscopic Eddington luminosity). Still, the fraction emitted in radio remains small.

Detection of a radio burst from a Galactic magnetar by CHIME and STARE2 collaborations in coincidence with high energy bursts \citep{2020arXiv200510324T,2020arXiv200510828B,2020arXiv200506335M,2020arXiv200511178R,2020arXiv200512164T}, and the  similarity  of it's properties to the Fast Radio Bursts (FRBs),  gives credence to the magnetar origin of FRBs.  The most compelling model, in our view, is the "Solar paradigm": generation of coherent radio emission during magnetospheric reconnection events \citep{2002ApJ...580L..65L,2013arXiv1307.4924P,2020arXiv200505093L} 

In this paper we develop a novel, magnetar-specific, model for the generation of coherent emission during magnetospheric reconnection events.  It relies on a process well studied in the context of laboratory plasma physics: free electron laser (FEL). Conceptually, many attempts to apply laboratory plasma process to astrophysical environments run into the ``no engineer on site"  problem: laboratory devices are often fine-tuned by engineers to produce the desired result; astrophysical lasers have to produce such fine-tuning naturally.  As we argue below, magnetar \mss\ may indeed naturally produce an analogue of the fine-tuned free electron laser.

 \section{Outline of the model: free electron laser in magnetar \mss}

The free electron lasers  \citep{1951JAP....22..527M,1971JAP....42.1906M,1976PhLA...59..187C,1977PhRvL..38..892D,1989PhFlB...1....3R,1991RvMP...63..949C}  are operational laboratory devices that   have  high efficiency of energy transfer from the kinetic energy of particles  to the coherent radiation.
Let us first briefly outline the principles of  FEL, in a regime relevant to \mss\ of magnetars. FEL  involves  a fast  relativistic electron beam propagating through periodically arranged magnets, the wiggler.   In the moving
reference frame of the beam, the wiggler magnetic field
Lorentz-transforms into a backward propagating transverse electromagnetic wave.  The wiggler's \Bf\ induces transverse oscillations of the
electrons, and, most importantly longitudinal oscillations that lead to the creation of periodic density enhancements due to the ponderomotive force of the wiggler's field. 
The collective  backscatter of the wiggler-produced transverse
wave by  density perturbations leads to coherent emission of the antenna-type. (Antenna mechanism, in contrast to plasma maser, implies that each electron emits independently, but externally imposed perturbation -- in this case the wiggler field - forces all the electrons emit in phase.) Most importantly this processes typically saturates at the level that a large fraction of the initial electron energy is converted into radiation.

Astrophysical applications of the FEL concept then require self-creation of the beam and the wiggler field. As we discuss below both conditions can be achieved during reconnection events in magnetar \mss. 
First, the wiggler field is naturally produced as a firehose instability of counter-streaming plasma components on closed field lines of twisted \ms\ of magnetars, \S \ref{firehose}. 
Second, reconnection events in the \ms\ launch mildly fast particle beams, with \Lf\  $\gamma_b \sim$ few  hundered. These beams propagating though the preexisting wiggler field produce FEL emission.  

Magnetic fields in magnetars can reach quantum critical fields. At the same time we are interested in the production of coherent emission in the GHz range. Thus,
 the FEL in    \mss\ of magnetars  will   operate in the somewhat unusual  regime  of ultra-strong guide field, when the cyclotron frequency  $\om_B =  e B_0/(m_e c) $ ($B_0$ is the guide field)  is much larger than the plasma frequency $\om_p$  and the radiation frequency  $\om$,   $\om_B \gg \om_p, \om$ \citep{1974PhFl...17..463M,1979PhFl...22.1089K,1980PhFl...23.2376F,2013PhRvS..16i0701G}. In what follows we reconsider FEL operating in such dominant guide field.

\section{Generation of a wiggler: pre-flare \ms\ is unstable to firehose instability} 
\label{firehose}

\subsection{Plasma flow in magnetar \mss}

Magnetars \citep{TD95,tlk} \citep[see ][for review]{2017ARA&A..55..261K} produce emission by dissipation of magnetic energy. Currents flowing in the non-potential \ms\ 
produce  emission in persistent state (corresponding to Anomalous X-ray pulsars, AXPs). Instabilities in the \ms\ can lead to the generation of burst (observed as Soft Gamma-Ray repeaters, SGRs), similar to the case of Solar flares \citep{2015MNRAS.447.1407L}.  Similarly to the Sun \citep{2010ARA&A..48..241B} radio emission can also be produced during X-ray flares  \citep{2002ApJ...580L..65L,2020arXiv200505093L}.

\cite{2013ApJ...777..114B} developed a model of particle flow in (quasi) stationary  states of magnetars. Briefly, the twist of the magnetic field lines frozen in the conducting crust
generates a  counter-streaming flow of $e^\pm$ pair.  
The pair densities are related to the overall twist of the \ms\  $\Delta \phi \leq 1$ and somewhat model-dependent plasma multiplicity  ${ \cal{ M}} \sim 100$ \citep{tlk,2013ApJ...777..114B}
  The pre-flare plasma density can then be parametrized  by the twist angle $\Delta \phi$ of the non-potential \Bf\ and pair multiplicity  ${ \cal{ M}}$ \citep{tlk,2013ApJ...777..114B}
   \ba &&
   n_p ={ \cal{ M}} (\Delta \phi) \frac{B}{4\pi e r} =  \kappa \frac{B}{4\pi e r} 
   \nn &&
    \kappa={ \cal{ M}} (\Delta \phi)
    \label{np}
   \ea
   where $r$ is the local radial coordinate.     
The pair components stream with respect to each other with $\gamma_p \sim 100$ \citep{2013ApJ...777..114B}.

Next we demonstrate that such plasma is unstable to the generation of intense \Alfven waves via  the firehose instability. 

\subsection{Firehose instability in magnetar \mss}
 
Let's assume that  magnetospheric plasma is composed of two  dense, oppositely charged  beams with equal densities $n_p$  and {\Lf}s $\gamma_p$. In the momentum rest frame, assuming charge neutrality, the dispersion relation for plasma (longitudinal) and transverse modes read \citep[][prob. 10.2]{1986islp.book.....M}
\ba && 
1-\frac{2 \omega _p^2 \left(k^2 v^2+\omega ^2\right)}{\gamma_p ^3 \left(\omega ^2-k^2
   v^2\right)^2}=0
   \nn &&
   \left( 1-\frac{k^2}{\omega ^2} -  \left( \frac{(\omega -k v)^2}{(\omega -k v)^2-\frac{\omega _B^2}{\gamma_p ^2}}+\frac{(\omega + k v
   )^2}{(\omega + k v )^2-\frac{\omega _B^2}{\gamma_p ^2}}  \right)  \frac{\omega _p^2}{\gamma_p  \omega ^2} \right)^2
-
 \nn &&
\left(\frac{\omega -k v}{(\omega -k v)^2-\frac{\omega _B^2}{\gamma_p ^2}}-\frac{\omega + k v
   }{(\omega + k v )^2-\frac{\omega _B^2}{\gamma_p ^2}}\right)^2 \, \frac{\omega _B^2 \omega _p^4}{\gamma_p ^4 \omega ^4} =0
\ea
(we set speed of light to unity).

The Langmuir plasma mode,
\be
\om_L^2 =k^2
   v^2+\left(1  \pm \sqrt{1+ \frac{  4 \gamma_p ^3 k^2 v^2}{\omega _p^2 } }\right) \frac{\omega _p^2}{\gamma_p ^3}
   \ee
   (interestingly one the branches  becomes subluminal for $k \geq \sqrt{  2(1+v^2) \gamma_p } \om_p$)
   shows two-stream instability for $k \leq \frac{\sqrt{3} \omega _p}{2 \gamma_p ^{3/2} v}$ with maximal growth rate
   \be
   \Gamma_L = \frac{\omega _p}{2 \gamma_p ^{3/2}}
   \ee

   We are more interested in the transverse  mode. In the limit $\om_B \gg \om_p, \om$  and small $k$ the dispersion becomes
   \be
 \om_t^2 =  k ^2\left(1 \pm \frac{v  \om_P^2}{ 2 k  \om_B}\right) 
   \ee
   which shows the firehose instability for 
   \be
   k \leq \frac{  2 v  \om_P^2}{  \om_B}
   \ee
   The maximal growth rate is 
   \ba &&
   k^\ast = \frac{   v  \om_P^2}{  \om_B}
   \nn &&
   \Gamma_{f} = \frac{   v  \om_P^2}{  \om_B}
   \ea

   We find for the growth rate and the wave number of the most unstable mode
   \ba && 
   \Gamma_f \approx  \kappa \frac{c}{r}
   \nn && 
    k^\ast \approx  \frac{ \kappa}{r}
    \label{Gamma}
    \ea
The growth rate and the wavelength of the most unstable firehose mode are independent of the local \Bf. This ensures that the model is applicable to a wide range of magnetars' \Bfs.

    Comparing the growth rate to the period of a \NS\ ${\Omega}$,
    \be
    \frac{  \Gamma_f }{\Omega} =   \kappa \frac{c}{ \Omega r}
    \ee
    The minimal required twist-times-multiplicity parameter $\kappa$ for magnetar-type periods  near the surface is small 
    \be
     \kappa_{min}  = \frac{\Omega r}{c}= 2 \times 10^{-4} P_s^{-1}   \left( \frac{r}{R_{NS}} \right)
     \ee
     where $P_s$ is a period of the \NS\ in seconds.  Hence we expect $\kappa \geq \kappa_{min}$ and the development of the instability.
     
In conclusion,  we expect that  persistent  plasma flows  in magnetar \mss\ are  unstable to firehose instability.
     
\subsection{Saturation of the firehose  instability}

Non-resonant  firehose  instability excites long wavelength modes of \Bf\ oscillations (low frequency   \Alfven modes) with typical amplitude $\delta B$.  Particles propagating through the  wiggle  field remain at lowest Landau level for $\om_B \gg \gamma_p (  k^\ast c) $: no cyclotron emission is generated. 
 Particles propagating  along the curved \Bf\ lines  will   emit  curvature emission in the typical radius of curvature $R_c \sim (k^\ast) ^{-1} (B_0/ \delta B)$. It may be demonstrated that the energy of the     curvature   emission produced during growth time of instability is tiny.

Thus, the initial energy of the relative motion is spent mostly  on the generation of the fluctuating  \Bf\ $\delta B$:
\ba &&
2 n_p \gamma_p m_e c^2 \approx \frac{ \delta B ^2}{4 \pi}
\nn &&
\delta B = 4 \sqrt{\pi } c \sqrt{n_p} \sqrt{m_e} \sqrt{\gamma _p}= 5\times 10^5  b_q^{1/2} \gamma_p^{1/2} \kappa^{1/2}  \left( \frac{r}{R_{NS}} \right)^{-2}  {} {\rm Gauss}
\label{deltaB}
\ea
where $\lambda_C = \hbar/(m_e c)$ is the electron  Compton wavelength and we normalized \Bf\ to critical quantum field, $B_0 = b_q B_q$, $B_q= {c^3 m_e^2}/(e \hbar)$.
This is clearly an upper estimate on the intensity of the wiggler, as we 
 neglected possible losses.
 
We expect that the mode with the highest growth rate at $k^\ast$ will be dominant. This is an important assumption that needs to be verified via PIC simulations (Philippov priv. comm.)


\section{Free electron laser during reconnection events in magnetar \mss}

\subsection{Particle acceleration in reconnection events}
 \label{recon} 
 Particle acceleration during reconnection events lately  came to the forefront of high energy astrophysics. Particularly important was  the  observations of the Crab Nebula $\gamma_p$-ray flares  by Fermi and AGILE \cite{2011Sci...331..739A,2011Sci...331..736T,2012ApJ...749...26B}, that  in many ways are challenging our understanding of the importance of different particle acceleration mechanisms in astrophysical plasmas. These events offer tantalizing evidence in favor of relativistic reconnection \cite{2011ApJ...737L..40U,2012MNRAS.426.1374C,2017JPlPh..83f6302L,2017JPlPh..83f6301L,2018JPlPh..84b6301L} operating in astrophysical sources.

Spectra of particles accelerated in reconnection events, obtained via PIC simulations, show a large variety,  depending both on the plasma magnetization $\sigma$, and, importantly, overall configuration of the system (see below).  
First, in case of highly magnetized plasma with $\sigma \gg 1$ reconnection can produce very hard spectra $p\leq 2$ (where distribution function $f \propto \gamma_p^{-p}$, harder than the conventional limit $p\geq 2$   first-order Fermi acceleration at shocks \citep[\eg][non-linear effect can produce slightly smaller values]{1987PhR...154....1B}. 
Reconnection in highly magnetized plasmas can naturally produce hard spectra with spectral index approaching $p \sim 1$ in the limit of large magnetizations \cite{zenitani_01,guo_14,sironi_15,werner_17}.
Reconnection can then explain hard   radio spectral indices $\alpha \sim 0.1-0.2$ in Pulsar Wind Nebulae (PWNe) \cite{2014BASI...42...47G}, as argued by 
\cite{2018PhRvL.121y5101C,2019MNRAS.489.2403L,2020arXiv200506319L}.

Second, large scale properties of the plasma configuration can also affect the particles' spectrum.  On the one hand,  most of work on particle acceleration in relativistic reconnection events use the initial set-up 
in the so-called ``Harris equilibrium,'' with magnetic field lines reversing over a microscopic (skin-depth-thick) current layer imposed as initial condition \cite{zenitani_01,2014ApJ...783L..21S,guo_14,werner_16}. In these result the newly formed magnetic X-points are ``flat'': as a result fast X-point acceleration regime is subdominant to island mergers \citep{2018PhRvL.121y5101C}

To overcome this problem \cite{2017JPlPh..83f6301L,2017JPlPh..83f6302L,2018JPlPh..84b6301L} investigated particle acceleration during  {\it explosive relativistic reconnection}. In this case the  reconnection is driven by large-scale stresses, rather than microscopic plasma effects.
As a result, the collapsing X-Point have large opening angle: this leads to initial very efficient and fast acceleration of a few lucky particles to energies well beyond the  initial mean energy per particle. This fast accelerated stage is then followed by a slower acceleration during island mergers.

 Such a two-stage acceleration in reconnection  can also generate the wiggler field: more dense slower components are unstable toward firehose instability, creating a wiggler field for the fast beams; then the fast beam produces coherent  FEL emission.  Highly non-stationary plasma process is needed in this scenario, since  the wiggler needs to be generated before the high energy beams.

The magnetic energy per particle (the sigma parameter) in magnetar \mss\ evaluates to 
\be
\sigma = \frac{B^2}{ 4 \pi n m_e c^2} = \frac{R_{NS}}{\lambda_C} {b_q} \kappa^{-1}   \left( \frac{r}{R_{NS}} \right)^{-2}=
10^{16} {b_q} \kappa^{-1}   \left( \frac{r}{R_{NS}} \right)^{-2}
\label{sigma}
\ee
Modern PIC simulations doe not come close to (\ref{sigma}). This large $\sigma$ also implies that the real {\Lf}s will be limited by other processes, not the average magnetic energy per particle or the available potential.

\subsection{Kinematics of  FEL}

As discussed above,  the pre-flare state is composed of two counter-streaming plasma beams, that became firehose unstable and generated an effective wiggler field. Reconnection event in the \ms\ generates a fast beam propagating through this field. Next we consider beam dynamics in the wiggler dominated by the guide field. 
      FEL with guide case has been previously considered by  \citep{1974PhFl...17..463M,1979PhFl...22.1089K,1980PhFl...23.2376F,2013PhRvS..16i0701G}.    In   \mss\ of magnetars the  FEL operates in the somewhat unusual  regime  of ultra-dominant strong guide field $\om_B \gg \om_p, \om$. Below we re-derive the salient features of FEL  in this unusual regime.

 Consider  a fast beam of density $n_b$  propagating with  \Lf\ $\gamma_b$ in the combined guide  field $B_0$ and wiggler \Bf\ $\delta B$.   
In the frame of the fast electron beam with $\gamma_b \gg 1$ the wiggler field looks nearly as a transverse \EM\ wave with intensity $E'_w = \gamma_b \delta B$ and 
wavelength $ k^{\prime} \approx   \gamma_b k^{\ast}$. 

For a given beam \Lf\ $\gamma_b$ particles of the beam scatter the wiggler into high frequency EM mode. The double-boosted frequency of the wiggler
 \be
 \om = 4 \gamma_b^2 (k ^\ast c) = 4\gamma_b^2 \kappa \frac{c}{r}
 \label{drift}
 \ee
 As the beam propagates up in the \ms\ the emitted frequency decreases. This explains the frequency drifts observed in FRBs \citep{2019ApJ...876L..23H,2019Natur.566..235C,2019arXiv190803507T,2019arXiv190611305J}, as argued by  \cite{2020ApJ...889..135L}. 
 
Given the wiggler's wave vector  $k ^\ast$ and  the  observed frequency $\nu_{ob}$ we can estimate beam's  \Lf:
 \be
 \gamma _ b = \sqrt{ \frac{\nu_{ob} }{ 2 k^\ast c} } =  \sqrt{ \frac{\nu_{ob} r }{ 2  \kappa c} }= 130  \nu_9^{1/2} \kappa^{-1/2}  \left( \frac{r}{R_{NS}} \right)^{1/2},
 \label{gammab}
 \ee
a fairly mild \Lf\ by  pulsar standards ($\nu_9$ is the observed frequency in GHz).

\subsection{Magnetic undulator parameter}

   The  conventional wave undulator parameter is
   \be
   a = \frac{ e \delta B}{  k^\ast m c^2} 
   \ee   
   (in the absence of guide field this is a typical transverse momentum of the beam particles in units of $m_e c$).
    Using the  estimate of the fluctuating \Bf\ (\ref{deltaB}), the nonlinearity parameter $a$ becomes
 \be
 a = \left(  \frac{4 B e r \gamma _p}{c^2 \kappa  m_e} \right) ^{1/2} =
 2 \sqrt{\frac{r}{\Lambda_C}}\left(\frac{\gamma_p b_q}{\kappa} \right)^{1/2} =
  3 \times 10^8 \sqrt{ \frac{b_q \gamma_p }{\kappa} }  \left( \frac{r}{R_{NS}} \right)^{-1}  \gg 1
 \ee 
 Similarly to $\delta B$, this is an upper limit.
   
 
In fact, the wiggler parameter for  astrophysically-relevant case of guide-field dominated FEL is different. 
In the strongly guide-field dominated case we need to use magnetic  undulator parameter $a_B$ instead of  $a$  \citep{2017ApJ...838L..13L,2019arXiv190103260L}.

Briefly, in the gyration frame of a particle subject to strong circularly polarized \EM\   wave and the guide field,  the amplitude of velocity oscillation $v_\perp' $ follows from the equations of motion,
\be
a = v_\perp' \left( \frac{1}{1-v_\perp^{\prime,2}} + \frac{\om_{B_0}}{\om'} \right)
\ee
(there are no $z$-oscillation in circularly polarized wave of constant amplitude.) In the absence of the guide field, $p_\perp =  a$, while in the strongly guide-dominated case
 \ba &&
 v_\perp' \equiv a_B =  \frac{  \delta B}{B}=  2 \sqrt{\frac{\lambda_C}{r}} \left(\frac{\gamma_p \kappa}{b_q} \right)^{1/2} = 10^{-8} \left(\frac{\gamma_p \kappa}{b_q} \right)^{1/2}  \left( \frac{r}{R_{NS}} \right)^{-1/2} 
 \nn &&
 = 10^{-8} \kappa^{1/2} \gamma_p^{1/2} b_q^{-1/2} \left( \frac{r}{R_{NS}} \right)^{-1/2} \ll 1  
 \ea
 Parameter $a_B$ is the typical transverse momentum (in units of $m_e c$) of a particle subject to  strong \EM\ wave in the dominant guide \Bf.  Qualitatively,  in the presence of a guide field a particle accelerates in a field $\delta  E \sim  \delta B$ for time $\sim 1/\om_B$, not $\sim 1/\om$ as is the case of no guide field. In any astrophysically relevant case   $a_B \ll 1$.
 
 Since transverse \Bf\ changes during the Lorentz boost, while the parallel remains the same, the magnetic  undulator parameter $a_B$ is not frame invariant.  In the frame of the fast beam it
 is
 \be
 a_B'= \gamma_b a_B
 \ee

\subsection{Beam dynamics in guide-field-dominated wiggler}


Consider next beam dynamics in the field of linearly polarized wiggler field.  For highly relativistic beam we can neglect the difference between  beam velocity $\beta_b$ and unity: the wiggler is then nearly an electromagnetic wave in the frame of the fast beam. We find then (primes denote values measured in the frame of the beam)
\ba &&
\E_w' = E_w'  \sin \xi  {} {\bf e}_x
\nn &&
\B_w'=  E_w'  \sin \xi {} {\bf e}_y + B_0{} {\bf e}_z
\nn &&
  \xi = \om' ( t-z/\beta_b )
  \nn &&
  E_w' = \gamma_b \delta B
\ea

Since $p_\perp/(m_e c) \sim a_B \ll 1$ the transverse motion of particles induced by the wiggle is non-relativistic in the frame of the beam.
We find  for transverse components in the limit $v_z \ll 1$
\ba &&
v_y = \frac{\omega _B \omega _{\delta } \sin \xi }{\omega ^{\prime, 2}-\omega
   _B^2} \approx - \frac{ B_w'}{B} \sin \xi  = - a_B \sin \xi  
   \nn &&
   v_x= \frac{\omega'  \omega _{\delta } \cos \xi }{\omega  
   _B^2-\omega ^{\prime, 2}}\approx \frac{ B_w'}{B} \frac{\om'}{\om_B} \cos \xi = \frac{\om'}{\om_B}  a_B \cos  \xi  
   \nn &&
   \omega _{\delta } = \frac{  e B_w'}{m_e c}
   \label{vv}
      \ea
   Note that $v_y \gg v_x$: particles in the field of the wiggler move nearly linearly (this is an important fact for the resulting polarization.)
   
   The longitudinal component, also non-relativistic, 
\be
 v_z=      \frac{\omega _{\delta }^2 \cos ^2\xi }{2 \left(\omega _B^2-\omega ^{\prime, 2}\right)}=-\frac{\omega _{\delta }^2 (1+\cos (2 \xi ))}{4 \left(\omega ^{\prime, 2}-\omega _B^2\right)}
 \approx \left(\frac{ B_w'}{2B} \right)^2 (1+\cos (2 \xi ))= \frac{1+\cos (2 \xi )}{4} \gamma_b^2 a_B^2
  \label{vz}
\ee
oscillates at the double frequency of the wiggler in the beam's frame. 

\subsection{Ponderomotive density enhancements}

 The wiggler field induces transverse oscillation of the beam. In addition, gradients of the wiggler's intensity in the beam's frame will induce longitudinal oscillations  as we discuss next \citep[see][for more detailed discussion of particle trajectories in the combined fields of wiggler and guide field]{1982PhFl...25..736F}.
 
Given the axial  velocity (\ref{vz}) charge conservation of the beam's particles,
\be
\partial_t n_b + \partial_z ( v_z n_b)=0
\ee 
gives
\ba && 
n_b'(\xi) =\frac{2 \omega _{\delta }^2 n(\xi)  \sin (2 \xi )}{-4 ( \om^{\prime, 2} -\omega _B^2)+ (1+\cos (2 \xi )) \omega
   _{\delta }^2}
   \nn &&
   n_b'= n_{b,0}' \left(  \left(1-\frac{\omega _{\delta }^2}{4 ( \om^{\prime, 2} -\omega _B^2)  } \right)  -\frac{\omega _{\delta }^2 \cos (2 \xi )}{4 ( \om^{\prime, 2} -\omega _B^2) }  \right) 
   \approx 
    \nn &&
    n_{b,0}' \left(  1+ \frac{\omega _{\delta }^2 \cos (2 \xi )}{4 \omega _B^2} \right) \approx 
        n_{b,0}' \left(  1+ \frac{ B_w ^{\prime, 2}}{4B_0^2}   \cos (2 \xi )\right) = n_{b,0} \left(  1+ \gamma_b^2 \frac{a_B^2}{4}  \cos (2 \xi )\right)
        \label{deltaB11}
   \ea
   where $  n_{b,0}'$ is the average beam density in its frame. One clearly recognizes this as density perturbations induced by the longitudinal ponderomotive force of the  
 wiggler.

 These  charge oscillations  of the beam particles  should not be suppressed by Debye screening of the bulk plasma, hence  we need the frequency of  wiggler field in the beam's frame to be higher than the plasma frequency of the bulk plasma, measured in the beam's frame
 \ba  && 
 \gamma_b (k^\ast c)  \geq  \frac{\om_p}{( \gamma_b \gamma_p) ^{3/2}}
 \nn && 
 \gamma_b \geq \left( \frac{B e r}{m_e c^2 \kappa \gamma_p^3 } \right)^{1/5} = 2 \times 10^3 b_q^{1/5} \kappa^{-1/5}  \gamma_p^{-3/5}  \left( \frac{r}{R_{NS}} \right)^{-2/5} ,
 \ea
 a condition not too difficult to satisfy.

   Thus, the wiggler introduces periodic density fluctuations in the fast beam,  on the scale of the half  wiggler wavelength (measured in the beam's  rest frame).

    \subsection{Collective  scattering of wiggler field by density inhomogeneities in the beam}
   
   As the beam propagates through the  swiggler 
    each beam particle scatters the wiggler field.
    Qualitatively, each scattering cross-section  is enhanced by $1+a^2$,  but is suppressed by $(\om/\om_B)^2$ \citep{1975UsFiN.115..161Z}.  Since the velocity oscillations  are non-relativistic, Eq (\ref{vv}), the resulting cross-section  $\sigma_s$  for scattering is  the conventional $\sigma_s \sim \sigma_T (\om/\om_B)^2$ \citep[$\sigma_T$ is Thompson cross-section,][]{1976MNRAS.174...59B}: high \Bf\ suppresses electron-photon interaction.

The collective processes on the other hand can greatly amplify the wave-particles interactions. In particular, 
 density perturbations of the beam  (\ref{deltaB11}) oscillate in phase in the field of the wiggler. As a  result of coherent addition of emission of  $N$ partiles confined to regions smaller that the wavelength and  oscillating in phase, the final intensity scales as $N^2$: this is the principle of the antenna mechanism of   coherent emission. 
 
\begin{figure}[h!]
\includegraphics[width=.99\linewidth]{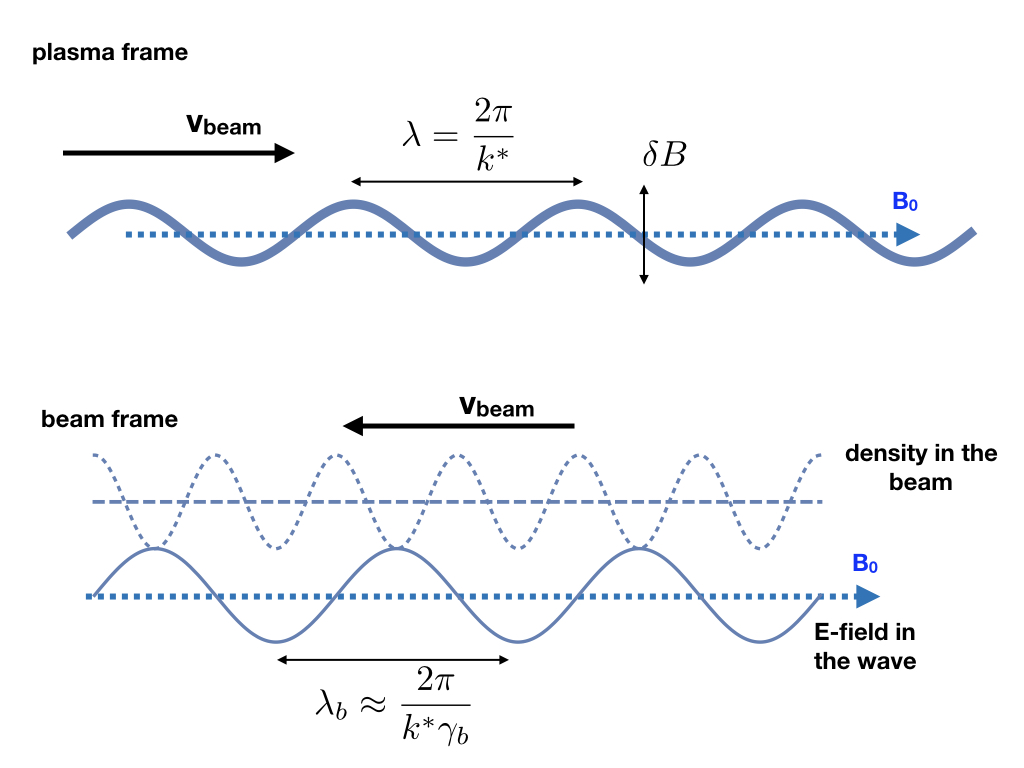}
\caption{Propagation of fast beam through  wiggler. In the wiggler frame the beam is propagating through perturbations with velocity $v_{beam}$. In the frame of the beam the wiggler is almost \EM\ wave that induces ponderomotive density perturbations at double the frequency of the wiggler in the frame of the beam. High density  regions of the beam oscillate in phase and coherently  scatter the wiggler's field.  }
\label{beam-density}
\end{figure}

 As we discussed above, the ponderomotive force of the wiggler creates density enhancements in the beam on scales smaller that the emitted (scattered) wavelength,  Eq. (\ref{deltaB11}) and Fig. \ref{beam-density}. These density enhancements oscillate coherently in the field of the wiggler, and  emit (scatter)  the wiggler's  radiation in phase.

The wiggler wavelength  in the frame of the beam is 
\be
\lambda^\ast_b = \frac{2\pi } {  \gamma_b k^\ast }
\ee

If the beam density in the lab frame is a fraction of the plasma density (\ref{np}),  $n_b = \eta _ b n_p$, beam density in the beam frame is
\be
n_b' = \frac{n_b}{\gamma_b} = \eta _ b  \frac{n_p}{\gamma_b}
\ee

Density enhancement in the beam frame   is
\be
\delta n'= n_b' \frac{ (\gamma_b \delta B)^2}{2 B_0^2} =\frac{  \eta _ b   \gamma_b   a_ B^2}{2}  n_p
\ee

Number of extra beam particles within the volume  $\lambda ^{\ast,3} _b$ is
\be
N_b' =  \lambda ^{\ast,3} _b \delta n'  =
{\gamma_p \eta_b} \frac{ m_e c^3}{\pi \nu_{ob}} = 10^{13} {\gamma_p \eta_b} \nu_9^{-1}
\label{nb}
\ee
(this is an upper estimate on $N_b' $ since density enhancement is limited just to a fraction of a wavelength).
The number $N_b' \gg 1 $ is the estimate of the  coherent enhancement of the single-particle scattering of the wiggler field in the dominant guide \Bf.

 Emissivity per particle in the beam's frame $P'$ and in  our frame $P$ are then
 \ba &&
 P' \approx \frac{e^2}{c}   \frac{ (\gamma_b \delta B)^2}{ B_0^2} N_b' 
 \nn &&
 P \approx 2  \gamma_b^2 P'= 32 \pi \gamma_b^2 \gamma_p \eta_b \kappa\frac{m_e c^3}{r}
 \label{Pco}
 \ea
 Using estimate (\ref{gammab}) for the \Lf\ of the beam we find a very simple relation for the coherent  power of each beam particle
 \be
  P \approx 16 \pi \eta_b \gamma_p \nu m_e c^2
  \label{Pp}
  \ee
 The coherent power per particle  (\ref{Pp}) is extraordinary high. For example, for beam-plasma equipartition $\gamma_b n_b \approx \gamma_p n_p \rightarrow 
 \eta_b = \gamma_p/\gamma_b$ relation (\ref{Pp}) implies that a beam particle loses it's energy to coherent emission in few oscillations. This is clearly an upper limit to the efficiency of coherent emission: recall that we used the upper limit on the wiggler's strength, Eq (\ref{deltaB}), and upper limit on the number of coherently emitting particles, Eq (\ref{nb}). Also dispersive effects of the beam, and finite bandwidth of the saturated wiggler modes were neglected. 

 As mentioned above, calculations of the saturation levels of coherent instabilities is an exceptionally complicated procedure.  Even with modern PIC methods very targeted types of simulations are needed to solidly assess the non-linear saturation level (since typically PICs use minimal resolution at plasma kinetic scales in an effort to capture larger scale dynamics.) Our  estimates of the  emitted power  are encouraging. 
 
\section{FEL and FRB/magnetar phenomenology}

The present model compares well with the observed FRB/magnetar phenomenology. The present model explains
\begin{itemize}
\item  contemporaneous radio-high energy flares.
\item high efficiency of conversion of particle energy into coherent radiation
\item  presence of narrow emission bands: it is related to the beam's \Lf\ and the wavelength of the wiggler field,  Eq. (\ref{drift})
\item variable emission properties from the same source \citep[\eg, two sub-bursts in][had somewhat different spectra]{2020arXiv200510324T}: mild variations of the beam \Lf, or of the frequency of the wiggler, lead to different emitted frequencies,  Eq. (\ref{drift}).
\item intermittency of radio production: specific combination of parameters of bulk plasma, and of the beam is required for observed emission to be produced, and to fall into the typical observational range of radio telescopes
\item  downward frequency drifts observed in FRBs,   \citep{2019ApJ...876L..23H,2019Natur.566..235C,2019arXiv190803507T,2019arXiv190611305J}. As the emission beam propagates in the magnetosphere the central frequency of the FEL decreases,  Eq. (\ref{drift}),  as argued  previously by  \cite{2020ApJ...889..135L}.
\item high linear polarization of FRBs  \citep[FRB 121102 and FRB 180916.J0158+65 show $\sim 100\%$ linear polarization,][] {2018Natur.553..182M,2019ApJ...885L..24C}.  In symmetric background pair plasma the wiggler is linearly polarized; in the particular FEL regime   the motion of beam particles is also nearly one-dimensional, Eq. (\ref{vv}).
\end{itemize}

The FEL likely operates during initial stages of magnetar flares
\citep[there is tantalizing evidence that radio bursts lead  X-ray bursts][]{2020arXiv200506335M}.
 It requires some minimal \Lf\ of the beam, Eq.  (\ref{gammab}). This may explain why only some, very hard,  X-ray flares \citep{2020arXiv200511178R} are accompanied by radio bursts: the coherence condition is not satisfied in most of the bursts, only in those that produce a beam with  sufficiently high  \Lf. 
 Another limitation is that the wave number of the unstable firehose mode should be sufficiently high, see Eq. (\ref{Gamma}). This requires high twist angles $\Delta \phi$ and high multiplicities ${\cal {M}}$.
    In some sense, fine-tuning of parameters is required to occasionally produce radio emission.  To address this question in more detail PIC simulations, including  pair production,  are needed.

\section{Discussion}
\label{Discussion}

In this paper we discuss a novel, magnetar-specific, model of  generation of coherent emission in \mss\ of \NSs: the free electron laser. 
We demonstrated first that the relative streaming of plasma component  in magnetar \mss\ is  firehose  unstable: this creates wiggler field that then scatters reconnection-produced fast beam.

The idea of a FEL in pulsar magnetosphere  has been previously briefly  discussed by \cite{SchopperFEL,FungKuijpers}. \cite{SchopperFEL} discuss FEL on plasma  Langmuir plasma turbulence generated by the two-stream Langmuir instability. Model of  Langmuir turbulence on the open magnetic  fields line of pulsar \mss\ run into problem of insufficiently high growth rate \citep{1977ApJ...212..800C,1986ATsir1431....1U}  \citep[see review by][]{2017RvMPP...1....5M}. On the open fields lines the plasma is moving with large bulk \Lf: this increases demands on the growth rate of the instability in the plasma frame. In addition, models based on the Langmuir waves excitation by the primary beam produced low growth rats due to the tenuous nature of the beam (with density only of the order of Goldreich-Julian), while 
 models based on the relative streaming of the secondary plasma faced a problem that for high density of the  secondary plasma  the relative velocity  develops slowly and remains small \citep{1987ApJ...320..333U}. 
 
Plasma flows in the magnetospheres of magnetars are different from the pulsars' open field lines: (i) plasma is not streaming with ultra-high \Lf\ along the \Bf\ (hence no suppression of instability due to bulk motion); (ii)   larger densities are expected on twisted \Bfs\ lines,   Eq. (\ref{np}); (iii) large relative velocity of the streaming components is expected \citep{2013ApJ...777..114B}. As a result, transverse  firehose perturbation are excited.

In terms of overall landscape of pulsar radio emission model, the 
present model falls under the ``antenna'' mechanism \citep[as opposed to plasma maser][]{1986islp.book.....M}:  all particles emit independently, but  in phase. In laboratories  
  this is achieved by the external driver  through frequency or amplitude modulation, or prearranged configuration of the wiggler field, while in the case  of magnetar \mss\ the ``driver'' is the results of the plasma instability in the pre-flare configuration.
   


\section*{Acknowledgements}

I would like to thank Andrei Beloborodov, Lorenzo Sironi and  Alexander Philippov for discussions. 
This research was supported by  NASA grant 80NSSC17K0757 and NSF grants 10001562 and 10001521.

\bibliographystyle{apj}

  \bibliography{/Users/maxim/Home/Research/BibTex}
  
  \end{document}